\documentclass[draft,a4paper,12pt]{article}
\usepackage{amssymb}
\title{Formal Properties of XML Grammars and Languages}
\author{Jean Berstel\\
\footnotesize{Institut Gaspard Monge (IGM)}\\[-6pt]
\footnotesize{Universit\'e de Marne-la-Vall\'ee}\\[-6pt]
\footnotesize{5, boulevard Descartes, 77454 Marne-la-Vall\'ee C\'{e}dex 2}\\
\and Luc Boasson\\
\footnotesize{Laboratoire d'informatique algorithmique: fondements  et
applications (LIAFA)}\\[-6pt]
\footnotesize{Universit\'e Denis-Diderot (Paris VII)}\\[-6pt]
\footnotesize{2, place Jussieu, 75251 Paris C\'{e}dex 05}
}
\begin{document}
\maketitle
\def\petitcarre{\vrule height4pt width 4pt depth0pt}
\def\QED{\relax\ifmmode\eqno{\hbox{\petitcarre}}\else
{\unskip\nobreak\hfil\penalty50
 \hskip2em\hbox{}\nobreak\hfil\petitcarre 
 \parfillskip=0pt \finalhyphendemerits=0\par\medskip\par}\fi}
\def\Proof{\medskip\par\noindent{\it Proof\/}}

\newtheorem{theorem}{Theorem}[section]
\newtheorem{proposition}[theorem]{Proposition}
\newtheorem{corollary}[theorem]{Corollary}
\newtheorem{lemma}[theorem]{Lemma}
\newtheorem{example}[theorem]{Example}
\newtheorem{remark}[theorem]{Remark}
\def\e{\varepsilon}
\def\a{\bar a}
\def\deriv{\mathop{\longrightarrow}\limits}
\def\Irr{\mathop{\rm Irr}\nolimits}

\begin{abstract}\noindent
XML documents are described by a document type definition (DTD). 
An XML-grammar is a formal 
grammar that captures the syntactic features of a DTD. 
We investigate properties of this family of grammars. We show that 
every XML-language basically has a unique
XML-grammar. We
give two
characterizations of languages generated by XML-grammars, one is
set-theoretic, the other is by a kind of saturation property. We 
investigate decidability problems and prove
that some properties that are undecidable for 
general context-free languages
become decidable for XML-languages. We also characterize those 
XML-grammars that generate regular XML-languages. 
\end{abstract}

\begin{quote}\begin{small}
\begin{center}\bf R\'esum\'e\end{center}
\noindent
Les documents XML sont d\'ecrits par une d\'efinition de type de document
(DTD). Une grammaire XML est une grammaire formelle qui retient les
aspects syntaxiques d'une DTD. Nous \'etudions les propri\'et\'es de cette
famille de grammaires. Nous montrons qu'un langage XML a
essentiellement une seule grammaire XML. Nous donnons deux
caract\'erisations des langages engendr\'es par les grammaires XML,
la premi\`ere est ensembliste, la deuxi\`eme est par une
propri\'et\'e de saturation. Nous examinons des probl\`emes de d\'ecision et
nous prouvons que certaines propri\'et\'es qui sont ind\'ecidables pour les
langages context-free g\'en\'eraux deviennent d\'ecidables pour les langages XML.
Nous caract\'erisons \'egalement les grammaires XML qui engendrent 
des langages rationnels.
\end{small}\end{quote}



\section{Introduction}

XML (eXtensible Markup Language) is a format recommended by W3C in order
to structure a document. The syntactic part of the
language describes the relative position of pairs of corresponding tags.
This description is by means of a document type definition (DTD). In
addition to its syntactic part, each tag may also have attributes.
If the attributes in the tags are ignored, 
a DTD appears to be a special kind of context-free grammar. The aim of this
paper is to study this family of grammars.

One of the consequences will be a better appraisal of the structure of
XML documents. It will also illustrate the kind of limitations that exist in the power of
expression of XML. Consider for instance an XML-document that consists
of a sequence of paragraphs. A first group of paragraphs is 
being typeset in bold,
a second one in italic. It is not possible to specify, by a DTD,
that in a valid document there are as many paragraphs in bold
than in italic. This is due to the fact that the context-free
grammars corresponding to DTD's are rather restricted. 

As another
example, assume that, in developing a DTD for mathematical documents, 
we require that in a (full) mathematical paper, there are as many 
proofs as there are statements, and moreover that proofs appear 
always after statements (in other words, the sequence of occurrences  
of statements and proofs is well-balanced). Again, there is no DTD for 
describing this kind of requirements. Pursuing in this direction, 
there is of course a strong analogy of pairs of tags in an XML 
document and the \verb=\begin{object}= and \verb=\end{object}= 
construction for environments in Latex. The Latex compiler merely 
checks that the constructs are well-formed, but there is no other 
structuring method.

The main results in this paper are two characterizations of
XML-lan\-gua\-ges. The first (Theorem~\ref{prop3}) is set-theoretic. It shows that XML-languages are
the biggest languages in some class of languages. It relies on the fact
that, for each XML-language, there is only one XML-grammar that
generates it. The second characterization (Theorem~\ref{prop1}) is
syntactic. It shows that XML-languages have a kind of ``saturation
property''. 

As usual, these results can be used to show that some languages cannot
be XML. This means in practice that, in order to achieve some features
of pages, additional nonsyntactic techniques have to be used.

The paper is organized as follows.
The next section contains the definition of XML-grammars and their
relation to DTD. Section~\ref{elementary} contains some elementary
results, and in particular the proof that there is a unique
XML-grammar for each XML-language. It appears that a new
concept plays an important role in XML-languages: the notion
of surface. The surface of an opening tag $a$ is the set of sequences of
opening tags that are children of $a$ (i. e. the tags immediately under
$a$ that may follow $a$ in a document
before the closing tag $\bar a$ is reached). The surfaces of an
XML-language must be regular sets, and in fact describe the
XML-grammar.
The characterization results are
given in Section~\ref{characterization}. They heavily rely on
surfaces, but the second also uses the syntactic concept of a context.

Section~\ref{decision}
investigates decision problems. It is shown that is is decidable
whether the language generated by a context-free language is
well-formed, but it is undecidable whether there is an XML-grammar
for it. On the contrary, it is decidable whether the surfaces of a
context-free grammar are finite (Section~\ref{FiniteSurfaces}). 

Section~\ref{regular} is concerned with regular XML-languages. It 
appears indeed that most XML-languages used in practical applications 
are regular. We show that, for a given regular language, it is 
decidable whether it is an XML-language, and we give a structural 
description of regular XML-grammars.

The final section is a historical
note. Indeed, several species of context-free grammars investigated in
the sixties, such as parenthesis grammars or bracketed grammars are
strongly related to XML-grammars. These relationships are sketched.

A preliminary version of this paper appears in the proceeding of the
MFCS~2000 conference \cite{bbmfcs}.

\section{Notation}

An XML document \cite{W3CXMLrec} is composed of text and of tags. The tags are opening or closing.
Each opening tag has a unique associated closing tag, and conversely.
There are also tags called empty tags, and which are both opening and closing. 
These tags may always be replaced by 
an opening tag immediately followed by its closing tag. We do so here,
and therefore assume that there are no empty tags. 

Let $A$ be a set of opening tags, and let $\bar A$ be the set of
 corresponding closing tags. Since we are interested in syntactic
 structure, we ignore any text. 
 Thus, an XML document (again with any attribute ignored) 
is a word over the alphabet $T=A\cup\bar A$.

A document $x$ is \textit{well-formed} if the word $x$ is a correctly
parenthesized word, that is if $x$ is in 
the set of Dyck primes over $A\cup\bar A$. Observe
that the word is a prime, so it is not a product of two well
parenthesized words. Also, it is not the empty word. 

An XML-\textit{grammar} is composed of a terminal alphabet $T=A\cup\bar A$,
of a
set of variables $V$ in one-to-one correspondence with $A$, of
a distinguished variable called the \textit{axiom} and, for each letter $a\in A$
of a regular set $R_a\subset V^*$ defining the (possibly infinite) set of productions
$$X_a\to am\bar a, \qquad m\in R_a,\quad a\in A$$ 
We also write for short 
$$X_a\to aR_a\bar a$$
as is done in DTD's.
An XML-\textit{language} is a language generated by some XML-grammar.

It is well-known from  formal language theory that
non-terminals in a context-free grammar may have infinite 
regular (or even context-free) sets
of productions, and that the generated language is still
context-free. Thus, any XML-language is context-free. Moreover, it is
a {\it deterministic} context-free language in the sense that there is 
a deterministic push-down automaton (\cite{Harrison}) recognizing 
it.

\begin{example} \rm
The language
$\{a^n\bar a^n\mid n>0\}$
is a XML-language, generated by
$$X\to a (X|\e) \bar a$$
\end{example}

\begin{example} \rm
The language of \textit{Dyck primes} over
$\{a,\bar a\}$ is a XML-language, generated by
$$X\to a X^* \bar a$$
\end{example}

\begin{example} \rm\label{E2}
The language $D_A$ of \textit{Dyck primes} over $T=A\cup\bar
A$ is generated by the grammar
$$\begin{array}{lcl}
X&\to&\sum_{a\in A} X_a\\[4pt]
X_a&\to& a X^* \bar a, \qquad a\in A
\end{array}$$
It is not an XML-language. However, each $X_a$ in this grammar generates
an XML-language, which is $D\cap aT\bar a$.
\end{example}

In the sequel, all grammars are assumed to be \textit{reduced}, that is, every non-terminal 
is accessible from the axiom, and every non-terminal produces at least
one terminal word. Note that for a regular (or even a recursive) set of
productions, the reduction procedure is effective.

Given a grammar $G$ over a terminal alphabet $T$ and a nonterminal $X$ we denote by
$$L_G(X) = \{w\in T^* \mid X \deriv^* w\}$$
the language generated by $X$ in the grammar $G$.

\begin{remark} \rm The definition has the following
correspondence to the terminology 
and notation used in the XML community (\cite{W3CXMLrec}).
The  grammar of a language is called a \textit{document type
definition} (DTD). The axiom of the grammar is 
qualified {\tt DOCTYPE}, and the set    
of productions associated to a tag is an {\tt ELEMENT}. The syntax of an
element implies by construction the one-to-one correspondence between
pairs of tags and non-terminals of the grammar. Indeed, an element
is composed of a \textit{type} and of a \textit{content model}. The type is
merely the tag name and the content model is a regular expression for the 
set of right-hand sides of the productions for this tag. For instance,
the grammar
$$\begin{array}{rcl}
S&\to&a(S|T)(S|T)\bar a\\
T&\to&bT^*\bar b
\end{array}$$
with axiom $S$ corresponds to
\begin{quote}
\begin{verbatim}
<!DOCTYPE a [
     <!ELEMENT a ((a|b),(a|b)) >
     <!ELEMENT b (b)* >
]>
\end{verbatim}
\end{quote}
Here, $S$ and $T$ stand for the nonterminals $X_a$ and $X_b$
respectively.

The regular expressions allowed for the content model are of two types:
those called children, and those called mixed \cite{W3CXMLrec}. In fact, since we do
not consider text, the mixed expressions are no more special
expressions.

In the definition of XML-grammars, we ignore {\it entities}, both general and
parameter entities. Indeed, these may be considered as shorthand and 
are handled at a lexical level.
\end{remark}

\begin{remark} \rm
    In the recent specification of XML Schemas (\cite{xmlSchema}), a 
    DTD is called a schema. The syntax used for defining schemas is 
    XML itself. Among the most significant enrichment of schema is the 
    use of types. Also the purely syntactical part of 
    XML schemas is more evolved than that of DTD's. 
\end{remark}

\section{Elementary Results}\label{elementary}

We denote by $D_a$ the language of
\textit{Dyck primes} starting with the letter $a$. This is the
language generated by $X_a$ in Example~\ref{E2}.
We set $D_A =\cup_{a\in A}D_a$. This is not an XML-language if $A$
has more than one letter. We call $D_A$ the set of Dyck primes over
$A$ and we omit the index $A$ if possible. The set $D$ is known to be a
\textit{bifix} code, that is no word in $D$ is a proper prefix or a
proper suffix of another word in $D$.

Let $L$ be any subset of the set $D$ of Dyck primes over $A$. The aim
of this section is to give a necessary and sufficient condition for
$L$ to be an XML-language.

We denote by $F(L)$ the set
of factors of $L$, and we set
$F_a(L) = D_a\cap F(L)$ for each letter $a\in A$.
Thus $F_a(L)$ is the set of those factors of words in $L$ that are
also Dyck primes starting with the letter $a$. These words are called 
{\it well-formed} factors.

\begin{example}\label{exb2n}
\rm
For the language
$$L=\{ab^{2n}\bar b^{2n}\bar a\mid n\ge1\}$$
one has $F_a(L)=L$ and $F_b(L) = \{b^n\bar b^n\mid n\ge1\}$.
\end{example}

\begin{example} \label{ex2}
\rm Consider the language
$$L=\{a(b\bar b)^n(c\bar c)^n\bar a\mid n\ge 1\}$$
Then $F_a(L)=L$, $F_b(L) = \{b\bar b\}$, $F_c(L)=\{c\bar c\}$.
\end{example}
The sets $F_a(L)$ are important for XML-languages and grammars,
as illustrated by the following lemma:

\begin{lemma}\label{lemme1}
Let $G$ be an XML-grammar over $A\cup \bar A$ generating a language $L$,
with nonterminals $X_a$, for $a\in A$. For each $a\in
A$, the language generated by $X_a$ is the set of factors
of words in $L$ that are Dyck primes starting with the letter $a$,
that is
$$L_G(X_a) = F_a(L)$$
\end{lemma}

\Proof.\ Set $T=A\cup \bar A$.
Consider first a word $w\in L_G(X_a)$. Clearly, $w$ is in
$D_a$. Moreover, since the grammar is reduced, there are words $g,d$
in $T^*$ such that $X\deriv^* gX_ad$, where $X$ is the axiom of $G$.
Thus $w$ is a factor of $L$.

Conversely, consider a word $w\in F_a(L)$ for some letter $a$,
let $g,d$ be a words such that $gwd\in L$. 
Due to the special form of an XML-grammar, any letter $a$ can only be
generated
by a production with non-terminal $X_a$. Thus,
a left derivation $X \deriv^* gwd$ factorizes into
\begin{equation}\label{eq1}
X \deriv^{k} gX_a\beta \deriv^* gwd
\end{equation}
for some word $\beta$, where $k$ is the number of letters in $g$ that
are in $A$. Next
\begin{equation}\label{eq2}
gX_a\beta \deriv^* gw'\beta \deriv^* gwd
\end{equation}
with $X_a \deriv^* w'$ and $w'\in D$. None of $w$ and $w'$ can be a
proper prefix of the other, because $D$ is bifix. Thus $w'=w$.
This shows that $w$ is in $L_G(X_a)$ and proves that $F_a = L_G(X_a)$.
\QED
\begin{corollary}\label{corollary}
For any  XML-language  $L\subset D_a$, one has $F_a(L)=L$. \QED
\end{corollary}

Let $w$ be a Dyck prime in $D_a$. It has a unique factorization
$$w=au_{a_1}u_{a_2}\cdots u_{a_n}\bar a$$
with $u_{a_i}\in D_{a_i}$ for $i=1,\ldots, n$. 
The \textit{trace} of
the word $w$ is defined to be the word $a_1a_2\cdots a_n\in A^*$.

If $L$ is any subset of $D$, and $w\in L$,
then the words $u_{a_i}$ are in $F_{a_i}(L)$.
The \textit{surface} of  
$a\in A$ in $L$ is the set $S_a(L)$ of all traces of words in $F_a(L)$.

\begin{example} \rm For the language of
Example~\ref{exb2n},
the surfaces are easily seen to be $S_a = \{b\}$ and $S_b = \{b,\varepsilon\}$.
\end{example}

\begin{example} \rm The surface of the language of Example~\ref{ex2}
are
$S_a = \{b^nc^n\mid n\ge 1\}$ and
$S_b=S_c=\{\e\}$.
\end{example}
It is easily seen that the surfaces of the set of Dyck primes over $A$
are all equal to $A^*$.

Surfaces are useful for defining XML-grammars. Let ${\cal S} = \{S_a\mid a\in A\}$ be a family of  regular
languages over $A$. 
We define an
XML-grammar $G$ associated to ${\cal S}$ 
called the {\it standard grammar} of ${\cal S}$  as follows. The 
set of variables is $V = \{X_a\mid a\in A\}$. For each letter $a$, we set
$$R_a = \{X_{a_1}X_{a_2}\cdots X_{a_n}\mid a_1a_2\cdots a_n\in S_a\}$$
and we define the productions to be
$$X_a \to am\bar a, \qquad m\in R_a$$
for all $a\in A$. Since $S_a$ is regular, the sets $R_a$ are regular over
the alphabet $V$. By construction, the surface of the language
generated by a variable $X_a$ is $S_a$, that is $S_a(L_G(X_a))=S_a$.
For any choice of the axiom, the
grammar is an XML-grammar. 

\begin{example}\label{exbn}\rm
 The standard grammar for the surfaces of Example~\ref{exb2n}
is 
$$\begin{array}{rcl}
X_a&\to&aX_b\bar a\\
X_b&\to&b(X_b|\varepsilon)\bar b
\end{array}$$
The language generated by $X_a$ is $\{ab^n\bar b^n\bar a\mid n\ge 1\}$ and
is {\it not} the language of Example~\ref{exb2n}.
\end{example}

This construction is in some sense the only
way to build  XML-grammars, as shown by the following proposition.

\begin{proposition}\label{prop3.5}
For each XML-language $L$, there exists exactly one reduced
XML-grammar generating $L$, up to renaming of the variables.
\end{proposition}

\Proof.\ Let $G$ be an XML-grammar generating $L$, with nonterminals
$V=\{X_a\mid a\in A\}$, and $R_a=\{m\in V^*\mid X_a\deriv am\bar a\}$
for each $a\in A$. We claim that
the mapping
$$X_{a_1}X_{a_2}\cdots X_{a_n} \mapsto a_1a_2\cdots a_n\eqno{(*)}$$
is a bijection from $R_a$ onto the surface $S_a(L)$ for each $a\in
A$. Since the surface depends only on the language, this suffices to
prove the proposition. It is clear that $(*)$ is a bijection from
$V^*$ onto $A^*$. It remains to show that its restriction to $R_a$ is
onto $S_a(L)$.

If 
$$X_a\deriv aX_{a_1}X_{a_2}\cdots X_{a_n}\bar a$$
is a production, then $a_1a_2\cdots a_n$ is the trace of some word $u$
in $L_G(X_a)$. By Lemma~\ref{lemme1}, the word $u$ is in $F_a(L)$, and
thus $a_1a_2\cdots a_n$ is in $S_a(L)$.

Conversely, if $a_1a_2\cdots a_n$ is in $S_a(L)$, then there is a word
$w\in F_a(L)=L_G(X_a)$ such that
$$w = a  u_1u_2\cdots u_n \bar a$$
with $u_i\in D_{a_i}$. Thus, there is a derivation
$$X_a\deriv am\bar a\deriv^* w$$
in $G$. Setting $m =Y_1Y_2\cdots Y_k$ with $Y_1,\ldots, Y_k\in V$,
there are words $u'_1,\ldots u'_k$ such that $Y_i\deriv^* u'_i$ and
$$u_1\cdots u_n=u'_1\cdots u'_k$$
However, each $u_i, u'_j$ is a Dyck prime, and since the sets of Dyck primes are
codes, it follows that $n=k$ and $u_i=u'_i$ for $i=1,\ldots, n$. Since
the words $u_i$ are in $F_{a_i}(L)$, there are derivations $X_{a_i}
\deriv^* u_i$. Thus $Y_i=X_{a_i}$ and 
$m=X_{a_1}X_{a_2}\cdots X_{a_n}$ as required. \QED

\begin{remark}\rm
Obviously, Proposition~\ref{prop3.5} is not longer true if entities are
allowed. Indeed, entities may be used to group sets of productions in
quite various manners.
\end{remark}

\begin{corollary}\label{corollary:inclusion}
Let $L_1$ and $L_2$ be two XML-languages. Then $L_1\subset L_2$ iff
$S_a(L_1) \subset S_a(L_2)$ for all $a$ in $A$. 
\end{corollary}
\Proof. The condition is
clearly necessary, and by the previous construction, it is also sufficient.
\QED

\begin{proposition}\label{inclusion}
 The inclusion and the equality of XML-languages is decidable.
\end{proposition}

\Proof. This follows directly from   Corollary~\ref{corollary:inclusion}. \QED

In particular, it is decidable if an XML-language $L$ is
empty. Similarly, it is decidable if $L=D_\alpha$.

XML-languages are not closed under union and
difference. This will be an easy example of the characterizations
given in the next section (Example~\ref{union}).

The following proposition is interesting from a practical point of view. 
Indeed, it shows that a stepwise refinement technique can be used in 
order to design a DTD that satisfies or at least approaches a given 
specification.

\begin{proposition} The intersection of two XML-languages is an XML-lan\-guage.
\end{proposition}

\Proof.\ Let $L$ and $L'$ be XML-languages generated by
XML-grammars $G$ and $G'$. We define an new grammar $G\times G'$with 
set of variables $V\times V'$ and productions
$$(X,X')\deriv a(X_1,X'_1)\cdots (X_n,X'_n)\bar a$$
if and only if $X\deriv aX_1\cdots X_n\bar a$ in $G$ and
$X'\deriv aX'_1\cdots X'_n\bar a$. The inclusion
$L_{G\times G'}(X,X')\subset L_{G}(X)\cap L_{G'}(X')$ is clear. Conversely,
assume $w\in L_{G}(X)\cap L_{G'}(X')$. Then
 $X\deriv aX_1\cdots X_n\bar
a\deriv^*w$ in $G$ and $X'\deriv aX'_1\cdots X'_{n'}\bar a\deriv^*w$
in $G'$. Thus $w=au_1\cdots u_n\bar a=au'_1\cdots u'_{n'}\bar a$,
where $X_i\deriv^*u_i$ and $X'_i\deriv^*u'_i$. Since the set of Dyck
primes is a code, one has $n=n'$ and $u_i=u'_i$. Thus $u_i\in
L_G(X_i)\cap L_G(X'_i)$ and the results follows by induction. \QED

\section{Two Characterizations of XML-languages}\label{characterization}

In this section, we give two characterizations of XML-language. The
first (Theorem~\ref{prop3}) is based on surfaces. It states that, for
a given set of regular surfaces, there is only one XML-language with
these surfaces, and that it is the maximal language in this family.
The second characterization   (Theorem~\ref{prop1}) is syntactical and
based on the notion of context. 

Let ${\cal S} = \{S_a \mid a\in A\}$, be a family of regular languages,
and fix a letter $a_0$ in $A$.
Define ${\cal
L}({\cal S})$ to be the family of languages $L\subset D_{a_0}$ such that $S_a(L)=S_a$
for all $a$ in $A$. Clearly, any union of sets in ${\cal L}({\cal S})$
is still in ${\cal L}({\cal S})$, so there is a maximal language (for
set inclusion) in this family. The {\it standard} language associated to 
${\cal S}$ is the language generated by $X_{a_0}$ in the standard grammar
of ${\cal S}$.

\begin{lemma}\label{lemme:maximal}
Let $L$ be the standard language of ${\cal S}$.
For any language $M$ in ${\cal
L}({\cal S})$, one has $F_{a_0}(M)\subset L$.
\end{lemma}
\Proof. Let $G$ be the standard grammar of ${\cal S}$. Then 
$L=L_G(X_{a_0})$.
We show that $F_a(M) \subset L_G(X_a)$ for $a\in A$
by induction on the length
of words. Let $w=au\bar a\in F_a(M)$. If $u$ is the empty word, then the
empty word is in $S_a$, and the word $a\a$ is in
$L_G(X_a)$. Otherwise,
$u$ has a (unique) factorization
$$u =u_{a_1}\cdots u_{a_n}$$
with $u_{a_i}\in F_{a_i}(M)$ for $i=1,\ldots, n$. By induction,
$u_{a_i}\in L_G(X_{a_i})$ for $i=1,\ldots, n$. Since
${a_1}\cdots{a_n}\in S_a$, there is a production
$X_a \to aX_{a_1}\cdots X_{a_n}\a$ in the grammar. Thus $w$ is in
$L_G(X_a)$. The result follows.
\QED

\begin{theorem}\label{prop3}
The standard language associated to
 ${\cal S}$ is the maximal element of the family ${\cal
L}({\cal S})$. This language is XML, and it is the only XML-language in
the family ${\cal L}({\cal S})$.
\end{theorem}

\Proof. The first part is just Lemma~\ref{lemme:maximal} and the
second part is Proposition~\ref{prop3.5}.
\QED

\begin{example} \rm The standard language associated to the sets
$S_a = \{b\}$ and $S_b = \{b,\varepsilon\}$
of Example~\ref{exb2n} is the language
$\{ab^n\bar b^n\bar a\mid n\ge 1\}$ 
of Example~\ref{exbn}. Thus, the language of Example~\ref{exb2n} is not XML.
\end{example}
We now give a more syntactic characterization of XML-languages. For this,
we define
the set of
 \textit{contexts} in $L$ of a word $w$ as the set $C_L(w)$ of pairs of words
$(x,y)$ such that $xwy\in L$.

\begin{theorem}\label{prop1}
A language $L$ over $A\cup\bar A$ is an XML-language if and only if 

{\rm (i)} $L\subset D_\alpha$ for some $\alpha\in A$,
 
{\rm (ii)} for all $a\in A$ and $w, w'\in F_a(L)$, one has
$C_L(w)=C_L(w')$,
 
{\rm (iii)} the set $S_a(L)$ is regular for all $a\in A$.
\end{theorem}

Before giving the proof, let us compute one example.
\begin{example}\label{exemple}\rm Consider the language $L$ generated by the
grammar
$$\begin{array}{rcl}
S &\to& aTT\bar a\\
T& \to& aTT\bar a\mid b\bar b
\end{array}$$ 
with axiom $S$. This grammar is not XML. Clearly, $L\subset D_a$. Also, $F_a(L)=L$. 
There is a unique set $C_L(w)$ for all $w\in L$, because at any place
in a word in $L$,
a factor $w$ in $L$ can be replaced by another factor $w'$ in
$L$. Finally,
$S_a(L)=(a\cup b)^2$ and $S_b(L)=\{\e\}$. The theorem claims that there is
an XML-grammar generating~$L$.
\end{example}

\Proof.\ We write $F_a$, $S_a$ and $C(w)$, with the language $L$ understood.
We first show that the conditions are sufficient. 

Let $G$ be the XML-grammar defined by the family $S_a$ and with axiom
$X_\alpha$. 
We prove first $L_G(X_a) = F_a$ for $a\in A$. 
By Lemma~\ref{lemme:maximal}, $F_a \subset L_G(X_a)$.
Next, we prove the inclusion $F_a \supset L_G(X_a)$ by induction on
the derivation length $k$. Assume $X_a \deriv^k w$. Then $w=au\a$ for
some word $u$. If $k=1$, then
the empty word is in $S_a$, which means that $a\a$ is in $F_a$. If
$k>1$,
then the derivation factorizes in
$$X_a \to aX_{a_1}\cdots X_{a_n}\a\ \deriv^{k-1} \ au\a$$
for some production $X_a \to aX_{a_1}\cdots X_{a_n}\bar a$. Thus there
is a factorization $u=u_1\cdots u_n$ such that $u_i\in L_G(X_{a_i})$
for $i=1,\ldots, n$. By induction, $u_i\in F_{a_i}$ for $i=1,\ldots, n$.
Moreover, the word $a_1\cdots a_n$ is in the surface $S_a$.
This means that there exist words $u'_i$ in $F_{a_i}$ such that the word
$w'=au'_1\cdots u'_n\a$ is in $F_a$. Let $g,d$ be two words such that
$gw'd$ is in the language $L$. Then the pair
$(ga,u'_2\cdots u'_n\a d)$ is a context for the word $u'_1$. By (ii),
it is also a context for $u_1$. Thus
$au_1u'_2\cdots u'_n\a$ is in $F_a$. Proceeding in this way, on strips
off all primes in the $u$'s, and eventually $au_1u_2\cdots u_n\a$ is
in $F_a$.
Thus $w$ is in $F_a$. This proves the inclusion and therefore the
equality. Finally, by Corollary~\ref{corollary}, on has
$L_G(X_\alpha)= L$, and consequently the conditions are sufficient.

We now show that the conditions are necessary. 
Let $G$ be an XML-grammar generating $L$,
with productions $X_a\to aR_a\a$ and axiom $X_\alpha$.
Clearly, $L$ is a subset of  $D_\alpha$. 
Next, consider words $w,w'\in F_a$ for some letter $a$, and
let $(g,d)$ be a context for $w$. Thus $gwd\in L$. By Lemma~\ref{lemme1}, we
know that $F_a = L_G(X_a)$. Thus, there exist derivations $X_a
\deriv^{*} w$ and $X_a \deriv^{*} w'$. Substituting the second to the
first in
\begin{equation}\label{eq3}
X_\alpha \deriv^* gX_ad\deriv^{*} gwd
\end{equation}
shows that $(g,d)$ is also a context for $w'$.
This proves condition (ii).

Finally, since $R_a$ is a regular set, the set $S_a$ is also regular.
\QED

\begin{example}\rm Consider the language $L$ 
of Example~\ref{exemple}. The construction of the proof of the theorem
gives the XML-grammar
$$\begin{array}{rcl}
X_a&\to& a(X_a|X_b)(X_a|X_b)\bar a\\
X_b&\to& b\bar b
\end{array}$$ 
\end{example}

\begin{example}\label{bolditalic} \rm The language
$$\{a(b\bar b)^n(c\bar c)^n\bar a\mid n\ge 1\}$$
already given above
is not XML since the surface of $a$ is the nonregular set $S_a = \{b^nc^n\mid
n\ge 1\}$. This is the formalization of the example given in the
introduction, if
the tag $b$ means bold paragraphs, and the tag $c$ means italic paragraphs.
\end{example}

\begin{example}\label{mathpaper} \rm In order to formalize the example of 
    well-formed mathematical papers given in the introduction, consider
the language
$L = \{aH\bar a\}$, where $H$ is the language obtained from the Dyck 
language over a single letter $b$ by replacing every $b$ by $t\bar t$ 
and every $\bar b$ by $p\bar p$. Here, the letters $t$ and $\bar t$ 
stand for \verb+<theorem>+ and \verb+</theorem>+ and $p$ and $\bar p$ 
for \verb+<proof>+ and \verb=</proof>= respectively. If one renames 
$t$ as $c$ and $p$ as $\bar c$, then the 
surface of $a$ in the language $L$ is the Dyck language over $c$, 
and it is not regular.
\end{example}

\begin{example} \rm Consider again the language
$$L=\{ab^{2n}\bar b^{2n}\bar a\mid n\ge1\}$$
of Example~\ref{exb2n}. First
$C_L(b\bar b) = \{(ab^{2n-1}, a\bar b^{2n-1}\bar a)\mid n\ge1\}$. Next
$C_L(b^2\bar b^2) = \{(ab^{2n}, a\bar b^{2n}\bar a)\mid n\ge0\}$.
Thus there are factors with distinct contexts. 
This shows again that the language is not XML.
\end{example}

Finally, we give an example showing that XML-languages are closed neither
under union nor under difference.
\begin{example}\label{union} \rm Consider the sets $cL\bar c$ and
$cM\bar c$, where $L=D_{\{a,b\}}^*$ is the set of products of Dyck
primes over $\{a,b\}$, and $M=D_{\{a,d\}}^*$ is the set of products of Dyck
primes over $\{a,d\}$. Each of these two languages is XML. However, the
union 
$H= L\cup M$ is not. Indeed, the words $cab\bar b \bar a\bar c$
and $ca\bar a d\bar d\bar c$ are both in $H$. The pair $(c, d\bar
d\bar c)$ is in the context of $a\bar a$, so it has to be in the
context of $ab\bar b \bar a$, but the word 
$c ab\bar b \bar ad\bar
d\bar c$ is not in $H$. 
Given a language $L\subset D_{a}$, write $\bar L = D_{a}-L$ for the 
relative complementation. Closure under difference would imply 
closure under relative complementation, and this would imply closure 
under union because $L\cup M =\overline{\bar L \cap \bar M}$. Thus
 XML-languages are not closed under difference.
\end{example}
 
\section{Decision problems}\label{decision}

As usual, we assume that
languages are given in an effective way, in general by a grammar or an
XML-grammar, according to the assumption of the statement. 

Some properties of XML-languages, such as inclusion or equality
(Proposition~\ref{inclusion}) are easily decidable because they reduce
to decidable properties of regular sets. 
The problem is different if one asks whether a context-free grammar
generates an XML-language. We have already seen in
Example~\ref{exemple} that there exist context-free grammars
that generate 
XML-languages without being XML-grammars.
We shall prove later (Proposition~\ref{prop:und}) that it is undecidable whether a context-free grammar
generates an XML-language. On the contrary, and in relation with Theorem~\ref{prop1},
it is interesting to note that it is decidable
whether a context-free language is a subset of the set of Dyck primes.
The following proposition and its proof
are an extension of a result by Knuth \cite{Knuth} who proved is for 
a single letter alphabet $A$.

\begin{proposition} \label{prop2} Given a context-free language $L$ over the
alphabet $A\cup \bar A$, it is decidable whether $L \subset D^{*}_A$.
\end{proposition}

We first introduce some notation. The \textit{Dyck reduction} is the
semi-Thue
reduction defined by the rules $a\a \to \e$ for $a\in A$. A word is
\textit{reduced}
or \textit{irreducible} if it cannot be further reduced, that means if
it has no factor of the form $a\a$. Every word $w$ reduces to a unique
irreducible word denoted $\rho(w)$. We also write $w\equiv w'$ when $\rho(w)=\rho(w')$.
If $w$ is a factor of some Dyck prime, then $\rho(w)$ has no factor of
the form $a\bar b$, for $a,b\in A$. Thus $\rho(w)\in \bar A^*A^*$. In
fact, $\rho(F(D_A))=\bar A^*A^*$.

\Proof\ of Proposition~\ref{prop2}.
Let $G = (V, P, S)$ be a (reduced) context-free grammar
(in the usual sense, that is with a finite number of productions)
 over $T= A\cup \bar A$,
with axiom $S\in V$, generating the language $L$. For each variable
 $X$, we set
$$ \Irr(X) = \{\rho(w) \mid X\deriv^* w, w\in T^*\}$$
This is the set of reduced words of all words generated by $X$. 
Testing whether $L$ is a subset of $D^{*}_A$ is equivalent to testing
whether $\Irr(S) = \{\e\}$.

First, we observe that if $\Irr(S) = \{\e\}$, then $\Irr(X)$ is finite
for each variable $X$. Indeed, consider any derivation $S\deriv^*
gXd$
with $g, d\in T^*$. Any $u\in \Irr(X)$ is of the form $u=\bar x
y$,
for $x,y\in A^*$.
Since $\rho(gud)=\rho(\rho(g) u \rho(d))=\e$,
the word $x$ is a suffix of $\rho(g)$, and $\bar y$ is a prefix of
$\rho(d)$. Thus $|u|\le |\rho(g)|+|\rho(d)|$, showing that the length of the
words in $\Irr(X)$ is bounded. This proves the claim.

A preliminary step in the decision procedure is to compute a candidate
to the upper bound on the length of words in  $\Irr(X)$. To do this,
one considers any derivation $S\deriv^* gXd\deriv^* gud$ with $gud\in
T^*$, and one computes $\ell_X = |\rho(g)|+|\rho(d)|$. As just mentioned
before, it is necessary that every reduced word in  $\Irr(X)$ has
length at most $\ell_X$.

We now inductively construct sets $\Irr_k(X)$ as follows. We start
with the sets $\Irr_0(X)=\emptyset$, for $X\in V$, and we obtain the sets in the
next step by substituting  irreducible sets of the current step in the
variables of the right-hand sides of productions. Formally, 
$$\Irr_{k+1}(X)= \Irr_k(X)\cup \bigcup_{X\to \alpha}
\rho(\sigma_k(\alpha))$$
where $\sigma_k$ is the substitution that replaces each variable $Y$
by the set $\Irr_k(Y)$. This construction is borrowed from 
\cite{Schutz}, with an addition use of the reduction map $\rho$ at each 
step. It follows that $\Irr(X) = \bigcup_{k\ge 0}\Irr_k(X)$

For each $k$, one computes $\Irr_k(X)$ for all $X\in V$, and then,
one checks whether $\Irr_k(X)= \Irr_{k-1}(X)$ for all $X$.
If so, the computation stops. The language $L$ is a subset of
$D_A$ if and only if $\Irr_k(S)=\{\e\}$.
If $\Irr_k(X')\ne \Irr_{k-1}(X')$ for some $X'$, then one checks
whether all words in $\Irr_k(X)$ have length smaller than
 $\ell_X$, for all $X$. If so, then one increases $k$. If the answer
is negative, then $L$ is not a
subset of $D_A$.

Since
the sets $\Irr_k(X)$ are finite, and the length of its elements must
be bounded by $\ell_X$ in order to continue, one eventually reaches a
step where the computation stops. \QED

\begin{corollary}\label{cor} Given a context-free language $L$ over the
alphabet $A\cup \bar A$ and a letter $a$ in $A$, it is decidable whether 
$L \subset D_a$.
\end{corollary}

\Proof. It is decidable whether $L\subset a(A\cup\bar A)^*\bar a$ (for 
instance by computing the set of first (last) letters of words in $L$. 
If this inclusion holds, then one effectively computes the language
$L'=a^{-1}L{\bar a}^{-1}$ obtained by removing the initial $a$ and the 
final $\bar a$ in all words of $L$. It follows by the structure of the 
Dyck set that $L\subset D_{a}$ if and only if $L'\subset D^*$. 
\QED

The proof of the following proposition uses standard arguments.

\begin{proposition}\label{prop:und} It is undecidable whether a context-free language
is an XML-language.
\end{proposition}
\Proof. Consider the Post Correspondence Problem (PCP) for two sets of words
$U=\{u_1,\ldots,u_n\}$ and $V=\{v_1,\ldots,v_n\}$ over the alphabet
$C=\{a,b\}$. Consider a new alphabet $B=\{a_1,\ldots,a_n\}$ and
define the sets $L_U$ and $L_V$ by
$$L_U=\{a_{i_1}\cdots a_{i_k}h\mid h\ne u_{i_k}\cdots
u_{i_1}\}
\quad
L_V=\{a_{i_1}\cdots a_{i_k}h\mid h\ne v_{i_k}\cdots
v_{i_1}\}$$
Recall that these are context-free, and that the set $L= L_U\cup L_V$
is regular iff 
$L=B^*C^*$. This holds iff the PCP has no solution. 

Set $A = \{a_1,\ldots,a_n, a,b,c\}$, and define a mapping $\hat w$ from
$A^*$ to $(A\cup\bar A)$ by mapping each letter $d$ to $d\bar d$.

Consider words $\hat u_1,\ldots, \hat  u_n, \hat v_1,\ldots, \hat v_n$ in $\{a\bar
a, b\bar b\}^+$
and consider the languages
$$\hat L_U=\{a_{i_1}\bar a_{i_1}\cdots a_{i_k}\bar a_{i_k}h
\mid h\ne \hat u_{i_k}\cdots
\hat u_{i_1}\}$$
and 
$$\hat L_V=\{a_{i_1}\bar a_{i_1}\cdots a_{i_k}\bar a_{i_k}h\mid h\ne \hat v_{i_k}\cdots
\hat v_{i_1}\}$$
Set $\hat L=c(\hat L_U\cup \hat L_V)\bar c $. The surface of $c$ in $\hat L$
is
$S_c(\hat L)= L_U\cup L_V$. If $\hat L$ is an XML-language, then
$L_U\cup L_V$
is regular which in turn implies that the PCP
has no solution. Conversely, if the PCP has no solution, $L_U\cup L_V$
is regular which implies that $L_U\cup L_V=B^*C^*$, which
implies that
$\hat L=c \hat B^* \hat C^*\hat c$, showing that $\hat L$ is an XML-language. 
\QED

\begin{corollary} Given a context-free subset of the Dyck set,
    it is undecidable whether its surfaces are regular. 
\end{corollary}

\Proof.\ With the notation of the proof of Proposition~\ref{prop:und},
the surface $S_c(\hat L)$ of the language $\hat L$ is the language
$L$,
and $L$ is regular iff the associated PCP has no solution. 
\QED

Despite the fact that regularity of surfaces is undecidable, it 
appears that finiteness of surfaces {\it is} decidable.
This is the main result of the next section.

\section{Finite Surfaces}\label{FiniteSurfaces}

There are several reasons to consider finite surfaces. First, the
associated XML-grammar is then a context-free grammar in the strict
sense, that is with a finite number of productions for each
nonterminal. 

Second, the question arises quite naturally within the
decidability area. Indeed, we have seen that it is undecidable whether
a context-free language is an XML-language. This is due basically to 
the fact that regularity of surfaces
is undecidable. On the other side, it \textit{is} decidable whether a context-free
language is contained in a Dyck language, and we will prove that it is
also decidable whether the surfaces are finite. So, the basic
undecidability result is the regularity of surfaces.

Finally, XML-grammars with finite surfaces are very close to families
of grammars that were studied a long time ago. They will be
considered in the concluding section.

\begin{theorem}\label{surfacesfinies} Given a context-free language $L$ that
is a subset of a set of Dyck primes, it is decidable whether $L$ has
all its surfaces finite.
\end{theorem}


\begin{corollary}\label{xmlsurffinies} Given a context-free language $L$ that
is a subset of a set of Dyck primes, it is decidable whether $L$ is 
a XML-language with finite surfaces.
\end{corollary}

In the rest of this section, we consider a reduced context-free grammar $G$
with nonterminal alphabet $V$, and terminal alphabet $T=A\cup \bar A$.
The language $L$ generated by $G$ is supposed to be a subset of some
set $D_\alpha$ of Dyck primes. Recall that $D = \bigcup_{a\in A}D_a$.
If $N$ is an integer such that $F(L)$ is contained in $D^{(N)}=\e\cup
D\cup D^2\cup\cdots\cup D^N$, we say that $L$ has {\it bounded width}.

First, observe that $L$ has finite surfaces iff it has bounded width.
Indeed, if the surface $S_a(L)$ is infinite for some $a\in A$, then
there are words of the form $au_1\cdots u_n\bar a$ in $F(L)$ for
infinitely many integers $n$, and clearly $F(L)$ is not contained in
any $D^{(N)}$. Conversely, if $u_1\cdots u_n\in F(L)$, then
there are words $w,w'\in D^*$ such that $awu_1\cdots u_nw'\bar a\in
F(L)$. But then the trace of this word has length at least $n$. Thus
if $F(L)$ is not contained in $D^{(N)}$, at least on surface is
infinite.

For the proof of the theorem, we investigate iterating pairs in $G$.
We start with a lemma of independent interest.

\begin{lemma}\label{lemmeX} If $X\deriv^+ gXd$ for some words in $g,d\in A\cup \bar
A)^*$, then there exist words $x,y,p,q\in A^*$ such that
$$\rho(g)=\bar xpx,\quad \rho(d)=\bar y\bar q y$$
and moreover $p$ and $q$ are conjugate words.
\end{lemma}
\Proof.\ The words $g$ and $d$ are factors of $D$. Thus, there exist
words $x,y,z,t\in A^*$ such that $g\equiv \bar xz $, $d \equiv \bar ty $.
There is a word $v$ such that $g^nvd^n$ is a factor of $D$
for each $n\ge 0$. From $g^2vd^2\equiv \bar x z\bar xz v
\bar ty\bar ty$, one gets that $x$ is a suffix of $z$ or $z$ is a
suffix of $x$, and similarly for $t$ and $y$. 
If $z$ is a suffix of $x$, set $x=pz$. But then $\bar z\bar p^n$ is a
prefix of  $\rho(g^nvd^n)$ for all $n$, contradicting the fact that
$\Irr(X)$ is finite. Thus $x$ is a suffix of $z$ and similarly $y$ is
a suffix of $t$. Set $z=px$ and $t=qy$. Then
$\rho(g)= \bar x p x$ and $\rho(d)=\bar y\bar q y$. Since 
$g^nvd^n\equiv \bar x p^n xv \bar y\bar q^n y$ and $\Irr(X)$ is finite,
one has $|p|=|q|$ and and moreover $p$ is a factor of $q^2$. \QED

A pair $(g,d)$ such that $X\deriv^+ gXd$ is a {\it lifting} pair if
the word $p$ in Lemma~\ref{lemmeX} is nonempty, it is a {\it flat}
pair if $p=\e$. 

\begin{lemma}\label{lemmeC} If $X\deriv^+ g_1Xd_1$ or $X\deriv^+ g_2Xd_2$ is a
lifting pair, then the compound pair $X\deriv^+ g_1g_2Xd_2d_1$
is a lifting pair.\end{lemma}
\Proof.\ According to Lemma~\ref{lemmeX}, $g_1\equiv \bar x_1p_1x_1$
and $g_2\equiv \bar x_2p_1x_2$. Assume the compound pair is flat. Then
$\bar x_1p_1x_1\bar x_2p_1x_2\equiv \bar z z$ for some word $z\in
A^*$. Thus the number of barred letters is the same as the number of
unbarred letters at both sides. This implies that $p_1$ and $p_2$ are
the empty word. \QED

\begin{lemma}\label{bounded} The language $L$ has bounded width iff $G$ has no
flat pair.
\end{lemma}
\Proof.\ If there is a flat pair $(g,d)$ in $G$, then $L$ has an
infinite surface. Indeed, $ug^nvd^nw\in L$ for all $n$ and for some
$u,v$,
and since $g\equiv \bar x x$, there is a conjugate of $g$ in $D$. Thus
$g^n$ has a factor in $D^{n-1}$, and $L$ has unbounded
width. 

Conversely, assume that $L$ has unbounded width. Let $K$ be the
maximum of the lengths of the right-hand sides of the productions in
$G$. Let $m$ be an integer that is strictly greater than the maximum of the
length  of the words in the (finite) sets $\Irr(X)$ for $X\in V$.
\def\Card{\mathop{\rm Card}}
Consider
a word $zu_1u_2\cdots u_Nz'\in L$ with $u_1,\ldots,u_N\in D$, for some
large integer $N$ to be fixed later. In a derivation tree for this
word,
let $X_0$ be the deepest node such that the tree rooted at $X_0$
generates a word containing the factor $u_1u_2\cdots u_N$. 
The production applied at that node has
the form $X\to Y_1\cdots Y_k$ with $Y_1,\ldots, Y_k\in V\cup T$ and
$k\le K$. By the pigeon-hole principle, at least one of $Y_1,\ldots,
Y_k$ generates a word containing a factor that is a product of at
least $N/k-1\ge N/K-1$ consecutive $u_i$'s. Denote this nonterminal
$X_1$. If $N$ is large enough, on constructs a sequence
$X_0,X_1,\ldots, X_h$ of nonterminals, and if $h\ge 
m\cdot\textrm{Card\,} V$, there are
at least $m$ of these variables that are the same. A straightforward
computation shows that $N\ge K+K^2+\cdots K^{m\cdot\,\Card V}$ is convenient.
We get pairs
$$\begin{array}{rcl}
Y&\deriv^*&s_1w_1p_1Yd_1\\
Y&\deriv^*&s_2w_2p_2Yd_2\\
&\cdots& \\
Y&\deriv^*&s_mw_mp_mYd_m\\
\end{array}$$
where each of $w_1,\ldots w_m$ is in $D^*$, the $s_i$ and $p_i$ are
suffixes (resp. prefixes) of words in $D$, and $p_1s_2, p_2s_3,\ldots
p_{m-1}s_m$ are Dyck primes. For each $i$, define $x_i\in A^*$ by setting $\bar x_i=\rho(s_i)$. From
$\rho(p_is_{i+1})=\e$, it follows that $\rho(p_i) = x_{i+1}$. Thus
$s_iw_ip_i\equiv \bar x_ix_{i+1}$. In view of Lemma~\ref{lemmeX},
there are words $y_i\in A^*$ such that $x_{i+1}=y_{i+1}x_i$ for
$i=1,\ldots m-1$, and each $s_iw_ip_i$ is equivalent to $\bar
x_iy_{i+1}x_i$, which in turn is equivalent to
$\bar x_1\bar y_2\cdots \bar y_{i} y_{i+1} y_{i}\cdots y_2 x_1$. All
$\bar x_1\bar y_2\cdots \bar y_{i}$ are prefixes of words in $\Irr(Y)$,
and since this set is finite, one of the $y_i$ is the empty word
because of the choice of $m$. This shows that one of the pairs is
flat. \QED

We now need to prove that it is decidable whether there exists a flat
pair. 
\begin{lemma}\label{lemmeI} 
Assume that $X\deriv^+ \ell_1Yr_1$,  $Y\deriv^+ gYd$ and $Y\deriv^+
\ell_2Xr_2$. If the pair $X\deriv^+ \ell_1g\ell_2Xr_2dr_1$ is flat,
then
the pair $Y\deriv^+ gYd$ is flat.
\end{lemma}
\Proof.\ According to Lemma~\ref{lemmeX}, $\ell_1g\ell_2\equiv \bar
zz$ and $g \equiv \bar x p x$ for some $z,x,p\in A^*$. 
Thus, $\ell_1g\ell_2$ has the same number of barred and of unbarred
letters,
and $g$ has more (or as many) unbarred letters than barred letters.
Next, 
$X\deriv^+\ell_1\ell_2Xr_2r_1$ is an iterating pair, and therefore
$\ell_1\ell_2$ has more unbarred letters than barred letters.
Thus $g$ has as many unbarred
letters than it has barred letters. It follows that $p$ is the empty word. \QED
\Proof\/ of Theorem~\ref{surfacesfinies}. In view of
Lemma~\ref{bounded}, it suffices to check whether the grammar has a
flat pair. For this, consider the derivation tree associated to a pair
$X\deriv^+ gXd$. We call this tree (and the pair) {\it elementary}
if there is no variable that is repeated on the path from the root $X$
to the leaf $X$. Lemmas~\ref{lemmeC} and \ref{lemmeI} shows that if
there is a flat pair, then there is also an elementary flat pair.

To each elementary pair, we associate a skeleton defined as
follow. Consider the path $X=X_0,X_1,\ldots, X_n=X$ from the root $X$
to the leaf $X$. Each of the $X_{i+1}$  is in the right-hand
side of some production $X_i\to \omega_i$. The skeleton is the
derivation obtained by composing these productions. It results in a
derivation $X\deriv^+ U X U'$, for some $U, U'\in (V\cup T)^*$. There
are only a finite number of skeletons because each skeleton is built
from an elementary pair. 

For each skeleton $X\deriv^+ U X U'$, we consider the set of pairs
$X\deriv^+ u X u'$ for all $u\in \Irr(U), u'\in \Irr(U')$ ($\Irr(U)$
denotes the set of reduced words of words deriving from $U$). Since all
$\Irr(U)$ is finite, the set of pairs obtained is finite. It suffices
to check whether there is a flat pair among them. \QED

As a final remark, we consider grammars
and languages similar to {\it parenthesis grammars} and languages 
studied by
Mc\-Naugh\-ton~\cite{McNaughton} and by Knuth~\cite{Knuth}. We will 
say more about them in Section~\ref{Historical}.
A {\it polyparenthesis grammar} is a grammar with a terminal alphabet
$T=A\cup \bar A$, and where every production is of the form
$X\deriv a m\bar a$, with $m\in V^*$, $a\in A$, $\bar a\in \bar A$. 
A polyparenthesis language is a language that has a polyparenthesis grammar.
Thus, polyparenthesis grammars differ from XML-grammars in two 
aspects: there are only finitely many productions, and the 
non-terminal need not to be unique for each pair $(a,\bar a)$ of 
letters.

\Proof\/ of Corollary~\ref{xmlsurffinies}. 
Let $G$ be a context-free grammar $G$ over
$A\cup \bar A$ generating $L=L(G)$. It is decidable whether 
$L\subset D_a$ for some letter $a\in A$ (Corollary~\ref{cor}). If this holds, we
check whether $L$ has finite surfaces. This is decidable 
(Theorem~\ref{surfacesfinies}).
If 
this holds, we proceed further. A generalization of an argument of 
Knuth \cite{Knuth} shows that it is decidable whether $L$ is a 
polyparenthesis language, and it is possible to effectively 
compute a polyparenthesis grammar 
$G'$ for it. On the other hand, let $G''$ be the standard grammar 
obtained from the (finite) surfaces. The language $L$ is XML if and 
only if $L=L(G'')$, thus if and only if $L(G')=L(G'')$. This equality 
is decidable. Indeed, any XML-grammar with finite set of productions 
is polyparenthetic, and equality of polyparenthesis grammars is  decidable
\cite{McNaughton}. \QED

\section{Regular XML languages}\label{regular}

Most of the XML languages encountered in practice are in fact 
regular. Therefore, it is interesting to investigate this case. The 
main result is that, contrary to the general case, it is decidable 
whether a regular language is XML. Moreover, XML-grammars generating 
regular languages will be shown to have a special form: they are 
{\it sequential} in the sense that its nonterminals can be ordered in such 
a way that the  nonterminal in the lefthand side of a production is always 
strictly less than the  nonterminals in the
righthand side. The main result of this section is 

\begin{theorem}\label{rat} Let $K\subset D_{A}$ be a regular language. 
    It is decidable whether $K$ is an XML-language.
\end{theorem}

One gets the following structure theorem.

\begin{proposition}\label{structure} Let $K$ be an XML-language, generated by an 
XML-gram\-mar $G$. Then $K$ is regular if and only if the grammar $G$ is 
sequential.
\end{proposition}

We shall give two proofs of Theorem~\ref{rat}, based on the two 
characterizations of XML-languages given above (Theorem~\ref{prop3} and 
Theorem~\ref{prop1}). Both proofs require the effective computation 
of surfaces.

\begin{lemma}\label{surfacelemma}
Let $K\subset D_{A}$ be a regular language. The surfaces of $K$ are 
effectively computable regular sets.
\end{lemma} 
\Proof.\ Let ${\cal A}$ be a finite automaton with no useless states
recognizing $K$. For each pair $(p,q)$ of states, let $K_{p,q}$ be
the regular language composed of the labels of paths starting in $p$ 
and ending in $q$. A pair $(p,q)$ of states is {\it good} for the 
letter $a$ in $A$, if $K_{p,q} \cap D_{a}\ne\emptyset$. This property is 
decidable. A pair is good if it is good for some letter. Let $G$ be 
the set of good pairs, considered as a new alphabet, and consider the
set $M(a)$ over $G$  composed of all words
$$(p_{0},p_{1})(p_{1},p_{2})\cdots (p_{n-1},p_{n})$$
such that there is an edge ending in $p_{0}$ in the automaton ${\cal A}$
and labeled by $a$ and there is an edge starting in $p_{n}$ labeled 
by $\bar a$. Clearly, $M(a)$ is a (local) regular language over $G$.

Consider now the finite substitution $f$ from $G^{*}$ into $A^{*}$ 
defined by
$$f(p,q)=\{a\in A\mid (p,q) \textrm{ is $a$-good}\}$$
Then $f(M(a))$ is the surface of $a$ in $K$, that is  $f(M(a))= 
S_{a}(K)$. This proves the lemma.
\QED

\medskip\par\noindent{\it First proof\/} of Theorem~\ref{rat}. We use 
Theorem~\ref{prop3}. Let $K$ be a regular subset of $D_{A}$. 
It is decidable whether $K\subset D_{a_{0}}$ for some letter $a_{0}$. 
If this holds, then by 
Lemma~\ref{surfacelemma}, the family ${\cal S}$ of surfaces $S_{a}(K)$
is effectively computable. From this family, one constructs the 
standard language $L$ associated to ${\cal S}$. This is effective. We 
know that $K\subset L$, and consequently $K$ is an XML-language if and 
only if $L \subset K$ or equivalently if and only if $L\cap K' = 
\emptyset$, where $K' = (A\cup \bar A)^*\setminus K$ is the 
complement of $K$. This is decidable. \QED

\medskip\par\noindent{\it Second proof\/} of Theorem~\ref{rat}. We use 
Theorem~\ref{prop1}.
 Let ${\cal A}$ be the minimal finite automaton with no useless states
recognizing $K$, with initial state $i$ and set of final states $T$. 
For each pair $(p,q)$ of states, let $K_{p,q}$ be
the regular language composed of the labels of paths starting in $p$ 
and ending in $q$. For each letter $a$ in $A$, 
the set $F_{a,p,q}=K_{p,q}\cap D_{a}$ is 
the set of well-formed factors of $K$ starting with the letter $a$ that are labels of 
paths from $p$ to $q$. Clearly, $F_{a,p,q}\subset F_{a}(K)$, for all 
$p,q$. We show that all words in $F_{a}(K)$ have same context if and 
only if $F_{a,p,q}= F_{a}(K)$, for all 
$p,q$ such that $F_{a,p,q}\ne\emptyset$. 

Assume first that all words in $F_{a}(K)$ have same context. Let 
$p,q$ such that $F_{a,p,q}\ne\emptyset$, and consider a word $w\in 
F_{a,p,q}$. There exist words $x$ and $y$ such that $i\cdot x = p$, 
and $q\cdot y\in T$. The pair $(x,y)$ is a context for $w$. Let $w'$ be 
a word in $F_{a}(K)$. Then there is a successful path with label 
$xw'y$. Thus there is a state $q'$ such that $p\cdot w' = q'$ and 
$q'\cdot y\in T$. If $q\ne q'$, there 
is a word $z$ separating $q$ and $q'$, because ${\cal A}$ is 
minimal. Thus $q\cdot z\in T$ and $q'\cdot z\notin T$ or 
vice-versa. However, this means that $(x,z)$ is a context for $w$ and 
is not a context for $w'$ or vice-versa. Thus $q=q'$ and $w'\in 
F_{a,p,q}$. This prove that $F_{a}(K)\subset F_{a,p,q}$.

Conversely, assume that $F_{a,p,q}= F_{a}(K)$, for all 
$p,q$ such that $F_{a,p,q}\ne\emptyset$. The contexts of any word
 $w\in F_{a}(K)$ is the union of sets 
$K_{i,p}\times K_{q,t}$ over all pairs $(p,q)$ with $F_{a,p,q}\ne\emptyset$.
Thus all words have same contexts.

It follows from the preceding claim that $K$ is a XML-language if and 
only if
$F_{a,p,q} = F_{a,p',q'}$ for all pairs for which the languages are
not empty. Although equality of context-free 
languages in not decidable in general, this particular equality is 
decidable because $F_{a, p,q}=F_{a, p',q'}$ iff
$$D_{a}\cap (K_{p,q}\setminus K_{p',q'}\cup K_{p',q'}\setminus 
K_{p,q}) = \emptyset \QED$$

For the proof of Proposition~\ref{structure} we use the following 
notation and result. For any word $w\in (A\cup\bar A)^*$, the {\it 
weight} of $w$ is the number $|w|_{A}-|w|_{\bar A}$. 
Here, $|u|_{A}$ is the number of occurrences of letters in $A$ in the 
word $u$.
The {\it height} of $w$ is the number
$$h(w) = \max\{|u|_{A}-|u|_{\bar A}\mid uv=w\}$$
that is the maximum of the weights of its prefixes.
The height of a language is the maximum of the 
heights of its words. This is finite or infinite.

\begin{proposition} Let $K\subset D_{A}$ be a
    language over $A\cup \bar A$. If $K$ is regular, then 
    it has finite height.
\end{proposition}

\Proof.\ This result is folklore. We just sketch its proof. Given an 
automaton recognizing $K$, the weight $|u|_{A}-|u|_{\bar A}$ of the 
label $u$ of a circuit must be zero for every circuit, by the pumping 
lemma. Thus, the height of $K$ is the maximum of the heights of the 
labels on all acyclic successful paths in the automaton augmented by 
the sum of the heights of all its simple cycles. Since the 
automaton is finite, this number is finite. \QED

\Proof\ of Proposition~\ref{structure}. Consider an XML-grammar $G$, 
and construct a graph with an edge $(X_{a}, X_{b})$ whenever $X_{b}$ 
appears in the righthand side of a production with $X_{a}$ as 
lefthand side. Nonterminals can be ordered to fulfill the condition of 
a sequential grammar if and only if the graph has no cycle. If the 
graph has no cycle, then the language generated by a variable of index 
$i$ is a regular expression of languages of higher indices. Thus, the 
language generated by the grammar $G$ is regular. On the contrary, if 
there is a cycle through some variable $X_{a}$, then there is a 
derivation of the form $X_{a}\deriv^*auX_{a}v\bar a$ for some words 
$u$, $v$. By iterating this derivation, one constructs words of 
arbitrary height in $K$, and so $K$ is not regular. \QED

Note that the language $F_{a}(K)$ of well-formed factors is regular 
when $K$ is a regular XML-language, because $F_{a}(K)$ is the language
generated by the nonterminal $X_{a}$ in a sequential grammar.

\section{Historical Note}\label{Historical}

There exist several families of context-free grammars  related
to XML-gram\-mars that have been studied in the past. In the sequel, the
alphabet of nonterminals is denoted by $V $.

\paragraph{Parenthesis grammars.} These grammars have been studied in
particular by
Mc\-Naugh\-ton~\cite{McNaughton} and by Knuth~\cite{Knuth}.
A {\it parenthesis grammar} is a grammar with terminal alphabet
$T=B\cup \{a,\bar a\}$, and where every production is of the form
$X\deriv a m \bar a$, with $m\in(B\cup V)^*$. A parenthesis grammar is
{\it pure} if $B=\emptyset$. In a parenthesis grammar, every
derivation step is marked, but there only one kind of tag.

\paragraph{Bracketed grammars.} These were investigated by Ginsburg
and Har\-ri\-son in \cite{Bracketed}. The terminal alphabet is of the form $T=A\cup \bar B\cup
C$
and productions are of the form $X\deriv a m \bar b$, with $m\in
(V\cup C)^*$. Moreover, there is a bijection between $A$ and the set
of productions. Thus, in a bracketed grammar, every derivation step is
marked, and the opening tag identify the production that is applied
(whereas in an XML-grammar they only give the nonterminal).

\paragraph{Very simple grammars.} These grammars were introduced by
Korenjak and Hopcroft~\cite{KorenjakHopcroft}, and studied in depth later on.
Here, the productions are of the form $X\deriv am$, with $a\in A$ and
$m\in V^*$. In a simple grammar, the pair $(a,m)$ determines the
production, and in a very simple grammar, there is only one production
for each $a$ in $A$.

\paragraph{Chomsky-Sch\"utzenberger grammars.} These grammars are used in the proof of the
Chomsky-Sch\"ut\-zen\-ber\-ger theorem (see e. g.~\cite{Harrison}),
even if they
were never studied for their own. Here the terminal alphabet is of the
form $T=A\cup \bar A \cup B$, and the productions are of the form
$X\deriv am\bar a$. Again, there is only one production for each
letter $a\in A$.\bigskip

XML-grammars differ from all these grammars by the fact that the set
of productions is not necessarily finite, but regular. However, one
could consider a common generalization, by introducing {\it balanced
grammars}.
In such a grammar, the terminal alphabet is $T=A\cup \bar A\cup B$,
and productions are of the form $X\deriv a m \bar a$, with $m\in
(V\cup B)^*$. Each of
the parenthesis grammars, bracketed grammars, Chomsky-Sch\"utzenberger
grammars are balanced. 
If
$B=\emptyset$, such a {\it pure} grammar covers XML-grammars with
finite surfaces. If the set of productions of each nonterminal is
allowed to be regular, one gets a new family of grammars with
interesting properties.

\end{document}